# Induced fit, conformational selection and independent dynamic segments: an extended view of binding events

**Peter Csermely**[1,a], **Robin Palotai**[1] and **Ruth Nussinov**[2,3]

[1]Department of Medical Chemistry, Semmelweis University, P O Box 260., H-1444 Budapest 8, Hungary
[2]Center for Cancer Research Nanobiology Program, SAIC-Frederick, Inc., NCI-Frederick, Frederick, MD 21702, USA
[3]Sackler Institute of Molecular Medicine, Department of Human Genetics and Molecular Medicine, Sackler School of Medicine, Tel Aviv University, Tel Aviv 69978, Israel

**Single molecule and NMR measurements of protein dynamics increasingly uncover the complexity of binding scenarios. Here we describe an extended conformational selection model which embraces a repertoire of selection and adjustment processes. Induced fit can be viewed as a subset of this repertoire, whose contribution is affected by the bond-types stabilizing the interaction and the differences between the interacting partners. We argue that protein segments whose dynamics are distinct from the rest of the protein ('discrete breathers') can govern conformational transitions and allosteric propagation that accompany binding processes, and as such may be more sensitive to mutational events. Additionally, we highlight the dynamic complexity of binding scenarios as they relate to events such as aggregation and signalling, and the crowded cellular environment.**

### The induced fit and the original conformational selection models
The prevailing view of binding mechanisms has evolved from the early 'lock-and-key' hypothesis [1] to the now popular 'induced fit' model [2]. According to the induced fit scenario, the interaction between a protein and a rigid binding partner induces a conformational change in the protein. However, several systems follow the 'conformational selection' (or, with a different nomenclature, population selection, fluctuation fit, selected fit) paradigm, where among the conformations of the dynamically fluctuating protein the ligand selects the one, which is compatible with binding, and shifts the conformational ensemble towards this state. Selective binding to a single conformation in the ensemble was suggested by Straub as early as 1964 [3]; initial experimental evidence for the hypothesis came from Zavodszky et al. [4] in 1966. Over twenty five years later, in a landmark paper, Frauenfelder, Sligar and Wolynes [5] described the energy landscape of proteins, which in 1999 led to the generalized concept of 'conformational selection and population shift' [6,7]. The observed widespread occurrence of such conformational selection phenomena and their importance in functional scenarios is increasingly supported by X-ray and cryo-electron microscope images, kinetics studies, extensive single molecule fluorescence and in particular, NMR data showing a repertoire of conformational states of unliganded proteins reflecting *in vivo* occurrences, including conformations corresponding to the bound form [8-10].

The duality of the induced fit and the original conformational selection models resembles the model pair of the Koshland-Nemethy-Filmer (KNF, [11]) and the Monod-Wyman-Changeux (MWC) models [12] which explain allosteric interactions. Both models describe the allosteric effect as a binding event at one site, which induces a conformational change affecting the activity at another site. However, the KNF-model views the conformational change as a consequence of allosteric binding, whereas the MWC-model describes the conformational change as an allosteric ligand-induced shift of the equilibrium of two, pre-existing conformational states. Recent data suggest that in several cases (e.g., adenylate kinase, catabolite activator protein) allostery can be mediated by transmitted changes in protein motions [13,14]. In this dynamic sense, allosteric regulation involves a population shift, that is, a change in the population of conformations preferring the state, whose binding site shape is complementary to the incoming partner. In many cases, the overall average conformation does not change, and the allosteric effect can only be seen by the increased (or decreased) dynamics of the protein segments [13,14].

---

[a]*Corresponding author:* Csermely, P. (csermely@eok.sote.hu)

Currently, it is well understood that the distinction between the induced fit and the original conformational selection models is not absolute; indeed, an increasing number of cases show that conformational selection is often followed by conformational adjustment [17,18]. The growing volume and precision of data relating to protein dynamics now enables a general discussion of binding events involving small ligands such as substrates, antigens or drugs, proteins and DNA and relating these to allosteric effects [6,8,19]. Extending the list of binding partners, Gorfe et al. [20] showed that conformational selection might play an important role in the insertion of proteins into membrane. Here, we use the above generality of binding scenarios to extend the original models, and to show that induced fit can be perceived as an extremity of this extended conformational selection model. We propose that independent dynamic segments, i.e. protein segments with distinct dynamics from the rest of the protein, could be key contributors to binding processes. Finally, we discuss intrinsically disordered proteins, extreme temperatures, aggregation, chaperones and crowded environment from the point of view of the extended conformational selection model, and show how this mechanism contributes to cellular signalling.

**The extended conformational selection model**

Recent data on protein dynamics makes the discrimination between the induced fit and the original conformational selection models less rigid. An increasing number of examples, in which conformational selection is followed by conformational adjustment [17,18], provide support for the extension of the original conformational selection model [6,7] (Figure 1). The extended conformational selection model describes the general scenario, where both selection- and adjustment-type steps follow each other. Recent data suggest that the conformational selection model also holds for RNA [21,22], which suggest that the extended conformational selection model could also describe the binding mechanism of RNA molecules.

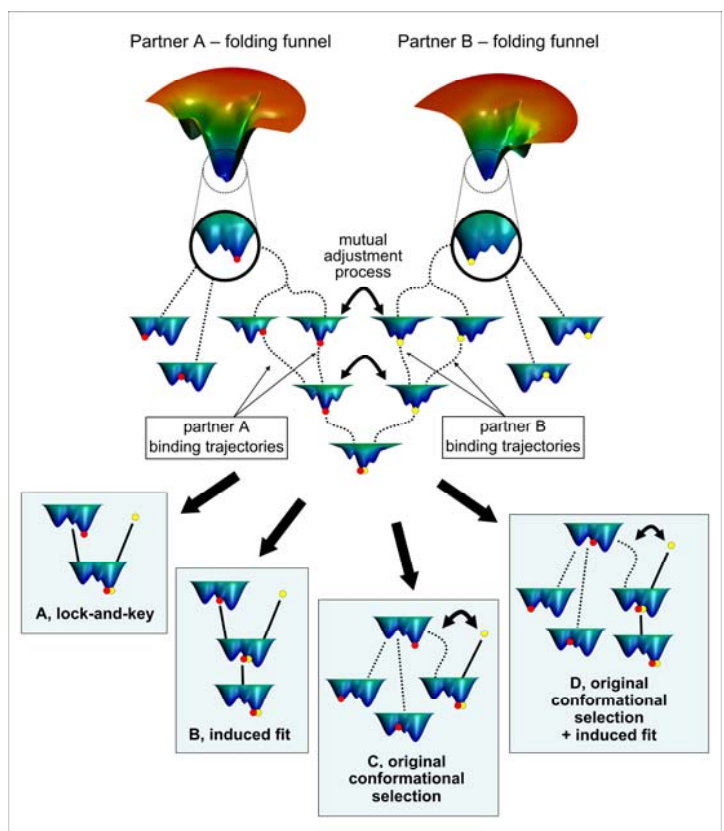

Figure 1. The extended conformational selection model. The 'native state' at the bottom of the two illustrative energy landscapes (folding funnels) of the two binding partners usually can be described as an assembly of conformers, which are visualized as secondary minima at the magnified bottom sections of the energy landscape in the circles and at the upper part of the figure. The top segment of the figure contains our proposed extended conformational selection model. For simplicity, we restrict our treatment to the encounter of two partners, because the probability for a simultaneous binding of more than two partners is extremely small. The actual conformation of partner A and B is described as their position on the energy landscape marked by red and yellow dots, respectively. Conformational selection is marked by the double headed arrows. As the two partners approach each other, the probability of occurrence of their conformers is changing, and so does the shape of their energy landscape. The actual sequence of these mutual conformational selection and preceding or subsequent conformational adjustment steps forms the 'binding trajectory' (marked with arrows) covered by partners A and B. For the sake of simplicity we restricted the number of binding trajectories to two. This number is several magnitudes higher in real situations. This general scenario is obviously much less complex, if the conformational ensemble of one of the partners is negligible compared to that of the other partner, e.g. in the case of binding of small substrates or allosteric modifiers to a protein as shown on the 4 panels. The lower part of the figure contains four scenarios of the extended conformational selection model, where binding trajectories are less complex. Panel A: the classical lock and key model, where both partners are either rigid, or have exactly matching binding surfaces. Panel B: the classical induced fit model, where Partner B first binds to the single available conformation of Partner A, and then induces a conformational change of Partner A. Panel C: the original conformational selection model, where Partner B binds one of the several fluctuating conformations of Partner A and no further conformational rearrangement occurs. Panel D: Partner B first binds one of the several fluctuating conformations of Partner A, and induces a subsequent conformational rearrangement of Partner B. As a discriminating feature of all scenarios of Panels A through D from the extended conformational selection model shown at the top centre, on all Panels Partner B has a single conformational status instead of a conformational ensemble. For the sake of simplicity its energy landscape is not shown. This approximation works well, if Partner B is small and/or rigid, like a small molecule or DNA. As a further simplification, Partner A's conformational landscape is not changed by Partner B on Panels A through D.

The ensemble of the conformational states available for mutual selection and adjustment is positioned at the low energy region of the folding funnel. Decades ago, this conformational ensemble was considered as the single conformational state of the 'native' protein. As binding proceeds not only do the partners' conformations change (accompanied by the changing position of the participating proteins on the energy landscape; Figure 1), but the mutual encounter also changes the shape of the energy landscape of both partners. As the two partners approach each other, electrostatic and water-mediated hydrogen-bonding signals emerge, and they increasingly change the partners' environment thereby altering their energy landscape [23,24]. Partner proteins can follow different sequences of conformational selection and adjustment steps (which we call 'binding trajectories' [13,24]). Such alternative pathways converging to a common end-state are also typical of the behaviour of other complex systems, such as gene expression of differentiating cells [25]. The encounter of the two binding partners involves a large number of conditional steps, where the next step of the encounter by partner A depends on a preceding conformational change by partner B and vice versa. We have previously compared this process to wooing, terming it an 'interdependent protein dance' [24]. Such a mutually conditional step-wise selection and encounter process can be described as a series of games ([23,24], Box 1).

---

**Box 1. The application of game theory models in the description of protein binding mechanisms.** In protein binding the conformation of one partner serves as an environment (a set of preconditions) for the other partner. This scenario is typical to a game, where the strategy (the repertoire of possible responses) of a partner depends on the last step of the other partner. The application of game theoretical models to protein binding was first proposed by Kovacs et al. [23], who suggested that binding events accompanied by parallel folding and unfolding, or unilateral folding (fly-casting) might correspond to well-defined games.

Protein binding might correspond to a hawk-dove game, where rigid proteins are hawks, whereas flexible proteins are doves [69]. In the hawk-dove game, if a hawk meets a hawk, a fight starts, both gain food but both are injured, and the cost of injury has to be deducted from the price of food at both sides symmetrically. If a dove meets a dove, the available food is shared equally. If a hawk meets a dove, the hawk will have the food, while the dove gets nothing. We propose to set the payoff of the hawk-dove game as the decrease of free energy. By this game definition, a rigid protein (a hawk) might indeed be a winner compared to a flexible protein (a dove), because the enthalpy gain of binding is not accompanied by an entropy cost for the rigid protein, but the more flexible protein loses several degrees of freedom during the binding event. If two rigid proteins (hawks) meet, no binding occurs, thus none of the partners wins anything. If two flexible proteins (doves) meet, the free energy-gain is shared. Induced fit corresponds to a hawk-dove encounter, while conformational selection corresponds to either a dove-dove or a hawk-dove game, where in the latter the 'protein-hawk' is selecting the appropriate conformation of the 'protein-dove'.

We also note that the *strictu senso* induced fit model resembles the ultimatum game. In this game the first player (the rigid protein) proposes how to divide the sum between the two players, and the second player (the flexible protein) can either accept or reject this proposal. If the second player rejects, neither player receives anything (i.e. no binding and no free energy decrease occurs). If the second player accepts, the money (the free energy decrease) is split according to the proposal. It is worth to mention that Chettaoui et al. [70] described a 'games network theory' to model multiple binding cascades as a network of different games and players. They efficiently applied this model to describe the signalling cascade of the plasminogen activator system containing 7 binding partners.

---

Our increasing knowledge of the details of the binding process might allow us to distinguish between the local and global versions of the original conformational selection and induced fit models [26]. As an example of this diversity, side chains in the ubiquitin binding site tend to follow induced fit type behaviour, whereas conformational selection type changes are more typical within the rest of the protein [18]. This distinction emphasizes that a unique categorization of the mechanism of the entire protein might not hold for all of its parts, thereby lending further support to our proposal of the extended conformational selection model described above (Figure 1).

## Special cases of the extended conformational selection model

The lock-and-key (Figure 1A), the induced fit (Figure 1B), the original conformational selection (Figure 1C) and the conformational selection + adjustment models (Figure 1D) are all special cases of the extended conformational selection model. In the lock-and-key model (Figure 1A) both partners have an exactly complementary binding surface. Conformational selection type binding scenarios (Figures 1C and 1D) shift towards the induced fit mechanism (Figure 1B), i.) if the interactions helping the mutual encounter are strong and long-range, like ionic interactions, or directed, like hydrogen-bond interactions [27]; ii.) if the partner's concentration is high [28,29]; and iii.) if there is a large difference in size or cooperativity [30]. These latter cases can be rationalized in terms of rigidity, as a small size ligand is often more rigid than its large protein partner which has a greater flexibility. Although ~ 90% of cellular protein–protein interactions are homomeric [30], thus precluding any of the above three scenarios leading to an induced fit, dimerization could involve different conformers of the same protein [31]. The duality of the induced fit (Figure 1B) and the original conformational selection models (Figure 1C) can also be rationalized in other complex systems (Box 2).

Box 2. The duality of the induced fit and the original conformational selection models in various complex systems. The induced fit and the original conformational selection models can be rationalized in complex systems beyond just macromolecules (Table I.). As the first example, the dipole–dipole interactions of two molecules can be viewed as an 'induced fit' or 'conformational selection' type interaction in the case of the induced dipoles (Debye forces) and fluctuating transient dipoles (London forces), respectively. Here the differences in the original 'rigidity' (an original, rigid dipole *versus* a molecule with a large polarization) help in the development of the 'induced fit' type Debye interactions from the London forces induced by fluctuating transient dipoles. Similarly, the pleiotropic nature of stem cells makes them ideally 'flexible' partners to adapt to the demands of their environment, which is a clear 'induced fit' type interaction, whereas, in general, neither of the interacting cortical neural cells can be regarded as much more flexible than the other making the establishment of inter-neuronal contacts similar to a 'conformational selection' process. Low diversity ecosystems, hierarchical animal or human communities (dictatures) are all determined by a few key species, which resembles an 'induced fit' type interaction, wheras the interactions in a high diversity system and in a cross-cooperating animal or human community (democracy) all resemble a 'conformational selection' model.

Table I. 'Induced fit' and 'conformational selection' in complex systems other than macromolecules

| Name of complex system member | 'induced fit' type interaction | 'conformational selection' type interaction |
| --- | --- | --- |
| Chemical molecules | induced dipoles (Debye-forces) | fluctuating transient dipoles (London-forces) |
| Cells | Stem cell differentiation | Multi-directional electronic and neurochemical interactions of cortical neurons |
| Ecosystem species | Food consumption in low diversity systems with a dominant herbivorous or carnivorous organism | Food consumption in high diversity systems with non-hierarchical (e.g. omnivorous) organisms |
| Animals | Division of labour in hierarchical communities of ants, bees, wasps, etc. having an leading figure (alpha male or female, queen) | Complementary cooperation between fishes, lions, monkeys, etc. without a dominant member of the community |
| Humans | Setting common norms and goals in a dictatorship | Setting common norms and goals in a democracy |

## Mechanisms of conformational dynamics during the binding process

The mechanism of the dynamic changes which occur during protein binding has been the subject of intense interest. The main question in these studies can be summarized as follows: Is the binding site the only key player, or are few selected regions in the protein involved? Over the past few years, an increasing number of key contributors to conformational changes have been identified. We discuss a few elementary steps of binding scenarios below, highlighting the importance of independent dynamic segments.

In the first step, transient encounter complexes are formed. These complexes were originally identified by paramagnetic relaxation enhancement experiments, are mostly stabilized by electrostatic forces, have a small, planar contact area in the range of only a few $Å^2$, and cover a rather large segment (e.g. 15%) of the total surface area around the binding site [31,32]. In the next step 'anchor residues' can play an important role. Anchor residues are in similar conformations as their final arrangement in the bound complex, and have a large surface area (~100 $Å^2$). Conformational selection of one to three anchor residues can be followed by an induced fit completion of the binding event involving several 'latch residues', which 'click in' to their final conformation, and thereby stabilize the interaction [34].

Anchor and latch residues are by far not the only protein segments involved in the conformational changes accompanying binding. Hinges and hinge-like motions were among the first suggested to play a decisive role in binding-induced conformational transitions [35]. Other studies revealed that critical nodes between communities of amino acid networks play an important role in the reorganization of the protein structure during a binding event ([36-39], Figure 2). Protein segments with 'discrete breather' behaviour accumulate kinetic energy and dynamically exchange as much as 20 to 65% of the accumulated energy during the conformational changes [40,41]. These segments are often located close to the binding or catalytic sites. We call these one or few amino acid long fragments 'independent dynamic segments' to emphasize their distinct behaviour. We note that hinges, critical nodes and independent dynamic segments might overlap, because independent dynamic segments are located in stiff, hinge-type regions of proteins, and both are often positioned at intermodular boundaries [35-38,40-42]. The possible involvement of specific regions of co-evolving amino acids, called 'protein sectors' has also been suggested [43]. 'Protein sectors' are sparse networks of amino acids, which span the entire protein, operate collectively and rather independently from each other. Segments of 'protein sectors', which are at the interface of protein domains, might constitute independent dynamic segments.

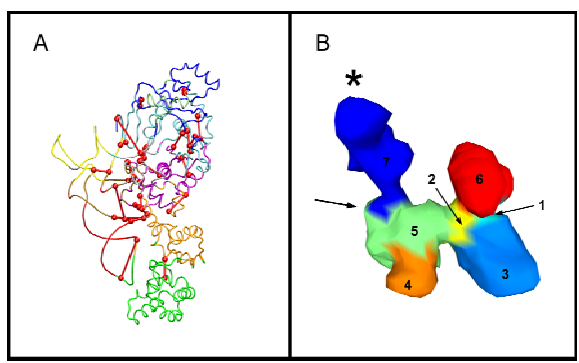

Figure 2. Inter-modular regions might act as key structural contributors of conformational changes in proteins. The two examples highlight amino acids located in the overlapping regions of residue-network modules. These inter-modular segments regulate the coordination of motion of the two modules they connect. Panel A: key inter-modular links and their end-point residues of Glu-tRNA synthase at the centre of the majority of shortest paths connecting nodes in different communities of the residue-network are marked with red lines and spheres [39]. Residue network communities are marked with different colours. (The image is a courtesy of Zaida Luthey-Shulten and John Eargle). Panel B: inter-modular regions of clusters of the ribosome-bound termination factor RF2 are marked with arrows. Clusters of correlated motions were determined from EMBD (Electron Microscopy Data Bank) data using normal mode analysis and were marked with different colours and numbers [66,67]. The asterisk denotes the peptidyl-transferase centre. The region between clusters 7 and 5, as well as clusters 1 and 2 connect clusters 7 (blue) and 6 (red) with the rest of the RF2 structure, respectively. Clusters 7 and 6 are the most distinctively moving segments of RF2 (The image is a courtesy of Mark Bathe and Do-Nyun Kim).

Independent dynamic segments might play a crucial role in binding. These independently vibrating protein segments can shift their energy content between each other [41]. This energy relocation might underlie the transition between various conformations in the conformational selection model and as such play a key role in allosteric propagation. Independent dynamic segments can also trigger induced-fit type mechanisms as in the conformational adjustment steps. In the relative absence of independent dynamic segments (due to high rigidity or flexibility) the protein probably lacks markedly differing conformational states, and its primary binding mechanism might shift from the original conformational selection model towards a lock-and-key or induced-fit type mechanism.

The binding process requires a gross reorganization of interactions (including desolvation), which is a source of frustration and conflict in need of efficient mediation [17,44-47]. Mediation of conflict can be provided by i.) key residues [44] which could be positioned at inter-modular segments and the independent dynamic segments described above [38,40,41]; ii.) transient bonds (e.g. transient, non-native hydrogen-bridges) [35,44,48]; iii.) water [25,44,45] and iv.) molecular chaperones [25]. We note that mediation is transient. Water molecules are expelled by the binding-induced, gradual desolvation [17,44-47], and molecular chaperones release their targets before complex formation is completed. Finally, independent dynamic segments [40,41] might become trapped as binding proceeds. By definition, before binding, independent dynamic segments have separate dynamics from the rest of the protein. Binding-induced stabilization might decrease the number and individuality of these key protein segments. In all cases, the decrease of mediation is gradual, and occurs in concert with the decreasing need for conflict mediation. This self-regulated withdrawal of conflict mediators, which provides their presence for optimal help, but disturbs the process at the possible minimum, is a true beauty of the complex events accompanying protein binding.

### Intertwined binding and folding events: intrinsically disordered proteins, extreme temperatures, aggregation and chaperones

Binding is often coupled with protein folding and unfolding. As an example of binding-induced protein folding, binding of tetracycline increases the stability and rigidity of the neighbouring DNA-binding domains, which ensures the correct positioning of the DNA-recognition helices of the two monomers in the major groove of the DNA [19]. Binding-induced folding is particularly evident in the case of intrinsically disordered proteins (IDPs). Importantly, a purely induced fit type binding mechanism is not characteristic even in these cases, because IDPs also have a broad fluctuating conformational ensemble, and conformational selection can take place [49-51]. The initial binding event of IDPs was proposed to be assisted by a 'fly-casting'-like search mechanism, where multiple weak binding events favour initial complex formation [52]. Subsequent, 'velcro'-like multivalent binding steps often increase the stability of the resulting complexes [49] illustrating further the utility of the proposed extended conformational selection model (Figure 1). A recent proposal posits that strongly interacting proteins can bind IDPs efficiently [53]. In these binding events, selection of a high-complementarity variant is followed by rigidification of the IDP partner. This rigidification step can be regarded as a subsequent induced fit, where the fit is achieved not by changing the average conformation, but by changing the distribution of the conformational sub-states in the assembly.

Temperature has a profound effect on protein stability and folding: cooling pauperizes, whereas heating expands the conformational ensemble. An increase in temperature makes the binding mechanism more similar to that of IDPs; however, at high temperatures both partners can become 'unstructured'. In nature efficient adaptation occurs: flexible loops or salt-bridges within proteins help to adapt organisms to extremely low or high temperatures, respectively [14]. At higher temperatures the contribution of ion-pairs and hydrogen bonds decreases, whereas that of hydrophobic bonds might increase. In comparison with other data [27], a lower involvement of strong, long-range and directed bonds in binding at higher temperatures may imply that the increase of temperature decreases the induced fit components of the binding mechanism, with a parallel increase in the conformational selection.

In recent years, aggregation has been increasingly perceived as a potential 'side-effect' of protein binding. Protein interface regions were proposed to be more prone to aggregation than other surface regions [54]. Aggregation is also a manifestation of conformational selection of higher energy (less populated) monomer states leading to highly polymorphic aggregate species.

Aggregation is often prevented by molecular chaperones, which can temporarily cover aggregation-sensitive surfaces. Chaperones also facilitate the assembly of protein complexes; indeed it was this property that led to the definition and name of this protein family [55]. Several types of molecular chaperones partially unfold their client proteins at the expense of their parallel folding [56]. Partial unfolding might promote the assembly of protein complexes, since one of the two binding partners become more similar to the IDPs. Chaperone-induced flexibility of one partner increases the possibility of an induced fit type scenario, and efficiently resolves the conflict of "who unfolds first", which often arises when two equally rigid partner proteins meet.

**Binding mechanisms in the crowded cellular environment**
Molecular crowding of the extremely packed cellular environment generally promotes the association of molecules *via* the excluded volume effect [55,57-59]. However, a crowded environment also decreases diffusion, which decreases the chances that two proteins will actually meet. Wang et al. [60] recently calculated the optimal intracellular concentration of proteins taking into account the balance of the opposing effects of the excluded volume effect-induced increased association and the decreased diffusion. They found that the optimal protein concentration is in the range of 1 mM, which is in the same range of the estimates of the actual intracellular protein concentration. The lower diffusion in crowded environments has a positive effect on binding in addition to decreasing the chances of protein encounter. Once the two binding partners find each other, crowding allows a longer contact time. The increased time in the vicinity of each other makes a more thorough conformational search possible, which might increase the dominance of conformational selection type processes in crowded environments.

**Conformational selection in signal transduction**
In signal transduction networks, the fine tuning of signalling targets and the requirement of alternative pathways ensuring the robustness of the response require conformational flexibility of the binding sites. Conformational fluctuations are not restricted to the nano- and picosecond range, but also occur in the range of milliseconds to seconds (or even hours), where they probably affect signalling steps [9,61]. This conformational dynamism opens the way for a large involvement of conformational selection in signalling-related binding events [62,63].

Recent data on the PDZ domains of the human tyrosine phosphatase 1E and calmodulin indicate that binding to a signalling protein at certain sites changes its conformational ensemble observed at another site. This induces a cascade of binding events providing both the exact targeting and amplification of the initial signal [29,62,63]. Phosphorylation and other posttranslational modifications involved in signalling shift the equilibrium of the initial conformational ensemble, and promote its subsequent binding by stabilizing a rarely populated pre-existing, active conformation of the protein [26,48]. Thus phosphorylation-induced stabilization of certain states provides a yet another example, whereby binding specificity is governed by folding-related events. A single step of conformational selection in its original sense generally helps to achieve a high specificity. However, if the selection process is not followed by a conformational rearrangement, then binding affinity could remain low. High affinity binding typically requires additional adjustment step; this step could be a general increase in protein dynamism at the binding site after specific binding has already been established [64] or conversely, a rigidification of the IDP partner [53]. However, other scenarios also exist: the small and flexible protein ubiquitin [8] undergoes a multitude of conformational selection steps to achieve sufficiently high affinity, while rigid spots at the binding site maintain the specificity.

The dynamic fluctuations of the conformational ensemble [13,29,62] can also be perceived as noise. Noise is not a havoc of efficient signalling; rather, it might help sigmoidal responses and switch-like behaviour (which can be perceived as an extremely steep sigmoidal response), which are crucial features of efficient signalling [9,61]. Increased noise might lead to the phenomenon of stochastic resonance [65], a process that occurs when noise occasionally helps a sub-threshold signal to surpass the sensitivity threshold. During signalling-related binding events, stochastic resonance might assist in populating certain relevant conformational states. Thus noise of protein dynamics appears to be an essential component of signal amplification and function.

## Concluding remarks and future perspectives

Recent data on protein dynamics uncovered the complexity of the mutual conformational selection and adjustment process and prompted us to suggest the extension of the original conformational selection model now including the classical lock-and-key, induced fit, conformational selection mechanisms and their combinations (Figure 1). The increasing precision of molecular dynamics simulations permits increasing mechanistic detail. Current observations implicate a key role of hinge-type regions, critical nodes at intermodular boundaries and in particular, independent dynamic segments in binding mechanism. Targeted mutations within these regions could modify the dynamics of the protein, that is, the distribution of the conformational ensemble and thus binding specificity. *In vivo*, such mutations can lead to disease.

With more data on protein dynamics and with the improvement of analytical tools, including network dynamics, perturbation analysis and game theory [23,24], further mechanistic details will be clarified. We believe that the identification of special protein regions governing binding mechanisms will be a major challenge in the near future; however, these advances would allow targeting of malfunctioning proteins in disease and in interface design. The analysis of the conformational dynamics of 452 of the 681 entries of the Electron Microscopy Data Bank (EMDB) [66,67] provides a novel rich source of data for the examination of protein dynamics.

Overall, instead of targeting entire proteins, specific interactions between proteins are becoming increasingly important targets for drug design [68]. Here we further propose that independent dynamic segments should be added to the allosteric drug design repertoire. We expect that the increasing knowledge of the dynamics and mechanism of binding processes we summarized in this paper would assist in developing drugs with substantially lower side effects and toxicity in the future.

**Acknowledgements**
The authors thank the Editor and the anonymous referees for their helpful suggestions, Mark Bathe and Do-Nyun Kim (Massachusetts Institute of Technology, Cambridge, MA USA), as well as Zaida Luthey-Shulten and John Eargle (University of Illinois at Urbana-Champaign, Urbana IL USA) for images of Figure 3. P.C. thanks Francesco Piazza (University of Portsmouth, UK), Yves-Henri Sanejouand (University of Nantes, France) and Gábor Simkó (Semmelweis University, Budapest, Hungary) for helpful discussions. Funding has been provided by the EU (FP6-518230) and by the Hungarian National Science Foundation (OTKA K69105). This project has been funded, in part, with federal funds from the National Cancer Institute, National Institutes of Health, under contract HHSN261200800001E. This research was supported, in part, by the Intramural Research Program of the NIH, National Cancer Institute, Center for Cancer Research. The content of this publication does not necessarily reflect the views or policies of the Department of Health and Human Services, nor does mention of trade names, commercial products, or organizations imply endorsement by the U.S. Government.